\documentclass[conference]{IEEEtran}
\IEEEoverridecommandlockouts
\usepackage{amsmath}
\usepackage{amsfonts}
\usepackage{bbding}
\usepackage{amssymb}
\usepackage{array}
\usepackage{subfigure}

\usepackage{graphicx}
\usepackage{subfigure}
\usepackage[named]{algo}
\usepackage{algorithmic}
\usepackage{algorithm}
\usepackage{psfrag}
\usepackage{xfrac}
\usepackage{stfloats}
\usepackage[compress]{cite}
\makeatletter
\renewcommand{\citepunct}{,\penalty\@m\hskip.13emplus.1emminus.1em}
\renewcommand{\citedash}{\hbox{--}\penalty\@m}
\makeatother
\usepackage{setspace}
\usepackage{color}
\allowdisplaybreaks

\usepackage{amsthm}
\usepackage{stfloats}

\newtheorem{rem}{Remark}

\usepackage{hyperref}
\usepackage{bm}

\begin{document}
\title{Unsupervised Deep Learning for Ultra-reliable and Low-latency Communications}

\author{
\IEEEauthorblockN{{Chengjian Sun and Chenyang Yang}} \vspace{0.0cm}
\IEEEauthorblockA{School of Electronics and Information Engineering,\\ Beihang University, Beijing, China\\
Email: \{sunchengjian,cyyang\}@buaa.edu.cn}\vspace{-0.6cm}
}
\maketitle
\begin{abstract}
In this paper, we study how to solve resource allocation problems in ultra-reliable and low-latency communications by unsupervised deep learning, which often yield functional optimization problems with quality-of-service (QoS) constraints. We take a joint power and bandwidth allocation problem as an example, which minimizes the total bandwidth required to guarantee the QoS of each user in terms of the delay bound and overall packet loss probability. The global optimal solution is found in a symmetric scenario. A neural network was introduced to find an approximated optimal solution in general scenarios, where the QoS is ensured by using the property that the optimal solution should satisfy as the ``supervision signal''. Simulation results show that the learning-based solution performs the same as the optimal solution in the symmetric scenario, and can save around $40\%$ bandwidth with respect to the state-of-the-art policy.

%In this work, we study how to reduce the required bandwidth to ensure the stringent quality-of-service (QoS) in ultra-reliable and low-latency communications (URLLC) by exploiting multi-user diversity. The resource allocation is in two timescales, where the power allocation is adaptive to small scale channels of multiple users and the bandwidth allocation is adaptive to their large scale channels. The joint power and bandwidth allocation is formulated as a functional optimization problem that minimizes the total bandwidth required to guarantee the QoS of all users. The global optimal solution is found in a symmetric scenario. An unsupervised deep learning methods, Deep Galerkin Method (DGM), is employed to find an approximated optimal solution in general scenarios. Numerical results demonstrate the mechanism of  multi-user diversity, which is to provide more resources to users with worse channel conditions. Simulation result show that the learning-based approximated solution performs the same as the optimal solution in the symmetric scenario, and can save around $40\%$ bandwidth in comparison to the state-of-the-art.
\end{abstract}

\section{Introduction}
Ultra-reliable and low-latency communications (URLLC) is one of the new application scenarios in the fifth generation cellular networks \cite{3GPP2016Scenarios}. Unprecedented quality-of-service (QoS) requirements on the end-to-end (E2E) latency (e.g., $1$~ms) and reliability (e.g., $10^{-5}$ packet loss probability) are demanded to support the mission-critical applications such as autonomous vehicles and smart factories \cite{Adnan2017Realizing}.

To improve the resource usage efficiency while ensuring the QoS of URLLC, various resource allocation problems have been investigated in the existing literature \cite{Changyang2017Mag, Schiessl2018ImperfectCSI, Samarakoon2018Federated, She2018CrossLayer, Chengjian2017GCw, Ben2018Risk, Elgabli2018AoI}. To ensure the packet error/loss probabilities and the queueing delay violation probability, the QoS constraint needs to be ensured for arbitrary large-scale channel gains. Since these probabilities rely on the resource allocation that should adapt to small-scale channel gains to ensure the short delay bound, the resource allocation problems involve two timescales, which are in fact functional optimization problems \cite{Zeidler2013Functional}.

In \cite{Samarakoon2018Federated}, power control was optimized to minimize the power consumption. To avoid the difficulty in solving the formulated problem, the original functional optimization problem is transformed into Lyapunov parameter optimization, which however is not equivalent to the original problem, and the resulting solution cannot satisfy the QoS requirements. In \cite{She2018CrossLayer, Chengjian2017GCw}, power and bandwidth allocation was investigated in multi-user scenarios. To avoid the difficulty in solving the formulated functional optimization problem, a power threshold was introduced to each user in \cite{She2018CrossLayer}. Then, the power threshold of and the bandwidth allocated to each user are optimized. Such a conservative design can ensure the QoS, but at the cost of using more resources.
In \cite{Chengjian2017GCw}, a heuristic resource allocation policy was proposed to take the advantage of  multi-user diversity. With multi-user diversity, the trade-off between reliability and resource usage efficiency can be improved, but the performance gap of the heuristic policy to the optimal solution is unknown. In \cite{Ben2018Risk, Elgabli2018AoI}, reinforcement learning was employed to solve the multi-timescale optimization problems in URLLC, where channel allocation and scheduling policies were learned according to the states of packet loss rate and the age of information, respectively. However, the reliability was controlled by taking the packet loss as penalties in the rewards, and hence the reliability cannot be ensured.

Functional optimization is challenging because it is the functions that need to be optimized, which can be interpreted as the vectors with infinite elements. Functional optimization problems usually do not has closed-form solutions, and need to be solved numerically, say by the Finite Element Method (FEM) \cite{Zienkiewicz1977FEM}.
%A typical functional optimization problem in wireless communication is to maximize the average service rate by allocating the transmit power over a fading channel with limited average transmit power. Water-filling algorithm can be adopted to find the analytical solution \cite{boyd}. However, the average service rate is insufficient to characterize the QoS requirements in URLLC and analytical solutions are not available for general functional optimization problems.
As a mesh-based method, FEM suffers from the curse of dimensionality. Resource allocation  in wireless systems is usually a multivariate function, e.g., the number of variables is equal to the number of users. When using FEM, the required number of elements increases exponentially with the number of variables of the functions, resulting in prohibitive computational complexity.
%
%The Deep Galerkin Method (DGM), which is an unsupervised deep learning method, was proposed in \cite{Sirignano2018DGM} to solve partial differential equations (PDEs). Since the functional optimization problems can be solved by finding the solutions of the KKT conditions, which are PDEs, the DGM is a promising candidate for solving the functional optimization problems. Moreover, without forming meshes, the DGM can circumvent the curse of dimensionality. However, \cite{Sirignano2018DGM} focused on how to satisfy the initial and boundary conditions in PDEs, how to ensure the constraints in the functional optimization problems is unclear.

In this paper, we study how to find the optimal solutions of functional optimization problems by resorting to unsupervised deep learning. We take downlink (DL) orthogonal frequency division
multiple access system supporting URLLC as an example. We optimize power and bandwidth allocation to minimize the required bandwidth to ensure the QoS of each user in URLLC by exploiting multi-user diversity. The QoS is characterized by the packet delay caused by DL transmission and queueing at the base station (BS) and the packet loss caused by decoding errors and queueing delay violation. We employ an accurate approximation of the achievable rate in short blocklength regime derived in \cite{Yury2014Quasi} to characterize the decoding error probability. We use effective capacity \cite{EC} and effective bandwidth \cite{EB} to control the queueing delay bound violation probability, which have been shown applicable for URLLC \cite{She2018CrossLayer}. The formulated problem needs functional optimization. To guarantee the QoS without using the costly labels for training the neural network, we use the property that the optimal solution should satisfy as the implicit ``supervision signal''. The basic idea is similar to the blind adaptive signal processing, say using the constant modulus property of communication signals as the label for training the policy \cite{CMA}.

%In this paper, we study how to fully exploit multi-user diversity to improve the spectrum efficiency in URLLC with the aid of neural networks while strictly guaranteeing the QoS requirements. To this end, we optimize the power control for each user as a function of the small-scale channel gains of all users. An accurate approximation of the achievable rate in short blocklength regime derived in \cite{Yury2014Quasi} is adopted to characterize the decoding error probability. Effective capacity \cite{EC} and effective bandwidth \cite{EB} are adopted to control the queueing delay bound violation probability. Since the delay distribution relies on the service process, which is related to the power control function, functional optimization cannot be avoided.
%Water-filling algorithm can be used to find the optimal power control that maximizes the average service rate in traditional communications \cite{She2015Context}, which however is not applicable to URLLC since the average service rate is insufficient to guarantee the delay requirement. In fact, for general cases, the optimal solution does not have closed-form expression. On the other hand, existing numerical methods (e.g., the Finite Element Method) are invalid for such a problem due to the curse of dimension.
%
%To solve such a challenging problem, the Deep Galerkin Method (DGM) is introduced to obtained numerical solutions, which is an unsupervised learning method that was first proposed in \cite{Sirignano2018DGM} to solve differential equations.

The major contributions are listed as follows.
\begin{itemize}
  \item In a symmetric scenario, we find the global optimal solution of joint power and bandwidth allocation.
  \item In general scenarios, we introduce an unsupervised deep learning to find an approximated optimal solution. We ensure the QoS by taking the Lagrange function of the problem as the loss function. Simulation results show that the learning-based solution performs the same as the optimal solution in the symmetric scenario, and saves around $40\%$ bandwidth compared to the heuristic policy in \cite{Chengjian2017GCw} in both symmetric and general scenarios.
\end{itemize}

The rest of the paper is organized as follows. In Section II, we introduce system model and define the QoS. In Section III, we formulate the resource allocation problem, show how to obtain the global optimal solution in symmetric scenario and to solve the problem in general scenarios with unsupervised deep learning. We provide simulation results in Section IV and conclude the work in Section V.

\section{System Model} \label{sec:system_model}
Consider a DL orthogonal frequency division multiple access system, where a BS with $N_\mathrm{t}$ antennas serves $K$ single-antenna users with maximal transmit power  $P_\mathrm{max}$. The bandwidth and the transmit power allocated to the $k$th user are $W_k$ and $P_k$, respectively.

Since the packet size $u$ in URLLC is typically small (e.g., $20$~bytes \cite{3GPP2016Scenarios}), the bandwidth required for transmitting each packet is less than the channel coherence bandwidth. Therefore, the channel is flat fading. Time is discretized into frames, each with duration $T_\mathrm{f}$. The duration for DL data transmission in one frame is $\tau$ and the duration for channel training is $T_\mathrm{f} \!-\! \tau$. Since the E2E delay requirement in URLLC is typically shorter than the channel coherence time, the channel is quasi-static and time diversity cannot be exploited. To guarantee the transmission reliability within the delay bound, we consider frequency hopping, where each user is assigned with different subchannels in adjacent frames. When the frequency interval between adjacent subchannels is larger than the coherence bandwidth, the small scale channel gains of a user among frames are mutual independent.

%\begin{figure}[htbp]
%	\vspace{-0.2cm}
%	\centering
%	\begin{minipage}[t]{0.48\textwidth}
%	\includegraphics[width=1\textwidth]{Resource_Block}
%	\end{minipage}
%	\vspace{-0.3cm}
%	\caption{Time and frequency resources allocation and frame structure.}
%	\label{fig:Frame}
%	\vspace{-0.2cm}
%\end{figure}

Packets desired by each user arrive at the buffer of the BS randomly. The inter-arrival time between packets could be shorter than the service time of each packet. Therefore, the packets may accumulate into a queue in the buffer. We consider a queueing model that the packets for different users wait in different queues.

\subsection{Achievable Rate in Finite Blocklength Regime}
In URLLC, the blocklength of channel coding is short due to the short transmission duration, and hence the impact of decoding errors on reliability cannot be ignored. Since Shannon's capacity formula cannot be employed to characterize the probability of decoding errors \cite{Gross2015Delay}, we consider the achievable rate in finite blocklength regime. In quasi-static flat fading channels, when channel state information is available at the transmitter and receiver, the achievable rate of the $k$th user (in packets/frame) can be accurately approximated by \cite{Yury2014Quasi},
\begin{align}   \label{eq:Srv}
    s_k \approx \frac{\tau W_k}{u \ln{2}} \left[\ln\left(1+\frac{\alpha_k g_k P_k}{ N_0 W_k}\right) - \sqrt{\frac{V_k}{\tau W_k}} Q_\mathrm{G}^{-1}\!\left({\varepsilon^\mathrm{c}_k}\right)\right],
\end{align}
where $\varepsilon^\mathrm{c}_k$ is the decoding error probability of the $k$th user, $\alpha_k$ and $g_k$ are the large-scale channel gain and small-scale channel gain of the $k$th user, respectively, $N_0$ is the single-side noise spectral density, $Q_\mathrm{G}^{-1} (x)$ is the inverse of the Gaussian Q-function, and $V_k$ is the channel dispersion given by \cite{Yury2014Quasi},
\begin{align}   \label{eq:Disp}
	V_k=1-\frac{1}{\left[1+\frac{\alpha_k g_k P_k}{N_0 W_k }\right]^2}.
\end{align}

Although the achievable rate is in closed-form, it is still too complicated to obtain graceful results. As shown in \cite{Gross2015Delay}, if the signal-to-noise ratio (SNR) $\frac{\alpha_k g_k P_k}{N_0 W_k } \ge$ $5$~dB, $V_k \!\approx\! 1$ is accurate. Since high SNR is required to ensure ultra-high reliability and ultra-low latency, such approximation is reasonable. Even when the SNR is not high, we can obtain a lower bound of the achievable rate by substituting $V_k \!\approx\! 1$ into $s_k$. Then, when the required $\varepsilon^\mathrm{c}$ is satisfied with the lower bound, it can also be satisfied with the achievable rate in \eqref{eq:Srv}.

\subsection{Quality-of-Service}
The QoS requirements of URLLC can be characterized by the delay bound $D_\mathrm{max}$ and the overall packet loss probability $\varepsilon_\mathrm{max}$. The uplink transmission delay, backhaul delay and processing delay have been studied in \cite{She2018Joint}, \cite{Gongzheng2016Backhaul} and \cite{Makki2018FastHARQ}, respectively, and are subtracted from the E2E delay in this paper. Thus, herein $D_\mathrm{max}$ is the DL delay, which consists of the queueing delay (denoted as $D^\mathrm{q}_k$ for the $k$th user), transmission delay $D^\mathrm{t}$ and decoding delay $D^\mathrm{c}$. All these delay components are measured in frames. $D^\mathrm{t}$ and $D^\mathrm{c}$ are constant values \cite{Condoluci2017Reserv}. Due to the random packet arrival, $D^\mathrm{q}_k$ is random. To ensure the delay requirement, $D^\mathrm{q}_k$ should be bounded by $D^\mathrm{q}_\mathrm{max} \!\triangleq\! D_\mathrm{max} \!-\! D^\mathrm{t} \!-\! D^\mathrm{c}$. If the queueing delay of a packet exceeds $D^q_\mathrm{max}$, the packet will be useless.
%To satisfy the QoS requirements in downlink, we need to ensure the probability that a packet can be successfully delivered within $D_\mathrm{max}$ to be higher than $1 \!-\! \varepsilon_\mathrm{max}$.

%\begin{figure}[htbp]
%	\vspace{-0.2cm}
%	\centering
%	\begin{minipage}[t]{0.3\textwidth}
%	\includegraphics[width=1\textwidth]{Delay}
%	\end{minipage}
%	\vspace{-0.3cm}
%	\caption{DL delay components.}
%	\label{fig:Delay}
%	\vspace{-0.2cm}
%\end{figure}

Denote $\varepsilon^\mathrm{q}_k \!\triangleq\! \Pr\{D^\mathrm{q}_k \!>\! D^\mathrm{q}_\mathrm{max}\}$ as the queueing delay violation probability.
Then, the overall reliability requirement can be characterized by
\begin{align}   \label{eq:Relia}
    1-(1-\varepsilon^\mathrm{c}_k)(1-\varepsilon^\mathrm{q}_k) \approx \varepsilon^\mathrm{c}_k+\varepsilon^\mathrm{q}_k \leq \varepsilon_\mathrm{max}.
\end{align}
This approximation is very accurate, because the values of $\varepsilon^\mathrm{c}$ and $\varepsilon^\mathrm{q}$ are very small in URLLC.

\section{Joint Power and Bandwidth Allocation}
 In this section, we show how to exploit multi-user diversity to minimize the total bandwidth required to support the QoS requirement in URLLC by optimizing resource allocation. We first obtain the global optimal solution in a special case, and then provide an approximated optimal solution for the general cases by resorting to unsupervised machine learning.

 \subsection{Problem Formulation and Equivalent Transformation}
 To exploit multi-user diversity, the transmit power allocated to each user is controlled according to the small-scale channel gains of all users $\bm{g} \!\triangleq\! (g_1,g_2,\cdots,g_K) \!\in\! \mathbb{R}_+^K$. In this way, the transmit power of the BS can be shared among users dynamically in each frame. Adaptively allocating bandwidth according to the small-scale channel
gains also yields multi-user diversity, which however can only bring marginal gain as demonstrated in \cite{Chengjian2017GCw}. To reduce the computational complexity, the bandwidth is only allocated to users according to their large-scale channel gains. Nonetheless, the method to be introduced in \ref{sec:DGM} is still applicable when the bandwidth allocation is adapted to $\bm{g}$.

 Since the transmit power depends on the small-scale channel gains, the packet service rate of each user is random. Further considering the randomness of the packet arrival, we use both effective capacity and effective bandwidth to analyze the queueing delay \cite{Lingjia2007TIT},\footnote{As analyzed in \cite{She2018CrossLayer}, if the frame
duration is much shorter than the delay bound, which is true in URLLC, effective bandwidth
can be used to analyze the queueing delay at the BS for Poisson, interrupted and switched Poisson arrival processes. We have validated that effective capacity can also be applied in URLLC, but do not show the results due to the space limitation.} with which the queueing delay violation probability of the $k$th user can be bounded by
 \begin{align}   \label{eq:QUpp}
    \varepsilon^\mathrm{q}_k < e^{-\theta_k B^\mathrm{E}_k D^\mathrm{q}_\mathrm{max}},
\end{align}
where $\theta_k$ is the QoS exponent that satisfies $C^\mathrm{E}_k \!\geq\! B^\mathrm{E}_k$, $C^\mathrm{E}_k$ and $B^\mathrm{E}_k$ are the effective capacity of the service process and the effective bandwidth of the arrival process of  the $k$th user, respectively.
Since the small-scale channel gains of a user are independent among frames owing to frequency hopping, the effective capacity of the $k$th user can be expressed as \cite{Tang2007TWC}
\begin{align}   \label{eq:EC}
    C^\mathrm{E}_k = - \frac{1}{\theta_k} \ln{\mathbb{E}_{\bm{g}} \left\{e^{- \theta_k s_k}\right\}} \; \text{(packets/frame)},
\end{align}
where the expectation is taken over the small-scale channel gains. Take the Poisson arrival process with the average packet arrival rate $a_k$~packets/frame as an example, whose effective bandwidth can be expressed as \cite{She2018CrossLayer}
\begin{align}   \label{eq:EB}
    B^\mathrm{E}_k = \frac{a_k}{\theta_k} \left(e^{\theta_k} - 1\right) \; \text{(packets/frame)}.
\end{align}

With the upper bound of $\varepsilon^\mathrm{q}_k$ in \eqref{eq:QUpp}, the queueing delay requirement ($D^\mathrm{q}_\mathrm{max}$,$\varepsilon^\mathrm{q}_k$) can be satisfied, and the overall reliability requirement in \eqref{eq:Relia} can be satisfied if
\begin{align}   \label{eq:ReliaCnsrv}
    \varepsilon^\mathrm{c}_k + e^{-\theta_k B^\mathrm{E}_k D^\mathrm{q}_\mathrm{max}} = \varepsilon_\mathrm{max}.
\end{align}
As shown in \cite{She2018CrossLayer}, the optimal values of the packet loss probabilities are in the same order of magnitude. Here we set $\varepsilon^\mathrm{c}_k \!=\! e^{-\theta_k B^\mathrm{E}_k D^\mathrm{q}_\mathrm{max}} \!=\! \varepsilon_\mathrm{max}/2$ for simplicity. Then, the QoS exponent corresponding to $D^\mathrm{q}_\mathrm{max}$ (and hence $D_\mathrm{max}$) and $\varepsilon_\mathrm{max}$ can be obtained from \eqref{eq:EB} as
$\theta_k = \ln \!{\left[1-\frac{\ln{(\varepsilon_\mathrm{max}/2)}}{a_k D^\mathrm{q}_\mathrm{max}}\right]}$.
If $C^\mathrm{E}_k$ in \eqref{eq:EC} is no less than $B^\mathrm{E}_k$ in \eqref{eq:EB} with $\theta_k$, the queueing delay requirement ($D^\mathrm{q}_\mathrm{max}$,$\varepsilon^\mathrm{q}_k$) with the upper bound of $\varepsilon^\mathrm{q}_k$ satisfying \eqref{eq:ReliaCnsrv} can be satisfied, and then the delay bound $D_\mathrm{max}$ and overall reliability $\varepsilon_\mathrm{max}$ can be satisfied.

The optimal power and bandwidth allocation problem that minimizes the total bandwidth required to ensure the QoS of every user can be formulated as,
\begin{align}   \label{prob:RA}
    \mathop \mathrm{min} \limits_{P_k(\bm{g}), W_k} \quad& \sum_{k=1}^{K} {W_k} \\
    \text{s.t.} \quad
    & - \frac{1}{\theta_k} \ln{\mathbb{E}_{\bm{g}} \left\{e^{- \theta_k s_k}\right\}} \geq B^\mathrm{E}_k, \label{con:Q} \tag{\theequation a} \\
    & s_k = \frac{\tau W_k}{u \ln{2}} \left[\ln\!\left(1+\frac{\alpha_k g_k P_k(\bm{g})}{ N_0 W_k}\right) -\frac{Q_\mathrm{G}^{-1}\!\left({\varepsilon^\mathrm{c}_k}\right)}{\sqrt{\tau W_k}}\right], \label{con:Srv} \tag{\theequation b} \\
    & \sum_{k=1}^{K} {P_k(\bm{g})} \leq P_\mathrm{max}, P_k(\bm{g}) \geq 0, W_k \geq 0, \label{con:Pmax} \tag{\theequation c}
\end{align}
where \eqref{con:Q} is the QoS requirement, \eqref{con:Srv} is the achievable packet rate in \eqref{eq:Srv} under the decoding reliability requirement with a power allocation function $P_k(\bm{g})$, and the first term in \eqref{con:Pmax} is the maximum transmit power constraint.

Problem \eqref{prob:RA} involves two timescales. The power allocation and bandwidth allocation adapt to the small-scale and large-scale channel gains, respectively. The queueing delay requirement should be satisfied for any large-scale channel gain (rather than for any small-scale channel gain). This makes the problem a functional optimization problem.

Moreover, the QoS constraint in \eqref{con:Q} does not have closed-form expression.
To solve such kind of problem, we can resort to stochastic optimization methods, such as stochastic gradient descent (SGD). To obtain an unbiased gradient estimation for SGD, the expectations in the objective function and constraints of a problem should not be in nonlinear forms. Thus, we transform \eqref{con:Q} into an equivalent form that is linear to the expectation, i.e.,
\begin{align}   \label{con:QLnr}
    \mathbb{E}_{\bm{g}} \left\{e^{- \theta_k s_k}\right\} - e^{-\theta_k B^\mathrm{E}_k} \leq 0.
\end{align}

Since less bandwidth is required if the queueing delay requirement is looser or more power resource is available, the optimal solution of problem \eqref{prob:RA} should be obtained when the equalities in \eqref{con:Q} and \eqref{con:Pmax} hold. Then, problem \eqref{prob:RA} can be equivalently transformed to the following problem,
\begin{align}   \label{prob:RALag}
    \mathop \mathrm{min} \limits_{P_k(\bm{g}), W_k} \mathop \mathrm{max} \limits_{h(\bm{g}), \lambda_k} \ & L \! \triangleq\! \sum_{k=1}^{K} {W_k} \!+\! \sum_{k=1}^{K} {\lambda_k \!\left( \mathbb{E}_{\bm{g}} \!\left\{\!e^{- \theta_k s_k}\!\right\} \!-\! e^{-\theta_k B^\mathrm{E}_k} \right)} \nonumber \\
    &\quad\  + \!\!\int_{\mathbb{R}_+^K} \!\!\!{h(\bm{g}) \!\left(\sum_{k=1}^{K} {P_k(\bm{g})} \!-\! P_\mathrm{max}\!\right) \!\mathrm{d}\bm{g}} \\
    \text{s.t.} \ & \eqref{con:Srv}, P_k(\bm{g}) \!\geq\! 0, W_k \!\geq\! 0, h(\bm{g}) \!\geq\! 0, \lambda_k \!\geq\! 0, \nonumber
\end{align}
where $L$ is the Lagrange function of problem \eqref{prob:RA}, and $h(\bm{g})$ and $\lambda_k$ are the Lagrange multipliers.

%Denote the probability density function of $\bm{g}$ as $f(\bm{g})$. Then, the optimal solution of problem \eqref{prob:RALag} should satisfy its Karush-Kuhn-Tucker (KKT) conditions, which can be derived as,
%\begin{align}
%    &\frac{\mathrm{\delta} L}{\mathrm{\delta} P_k(\bm{g})} = h(\bm{g}) - \lambda_k \theta_k \frac{\partial s_k}{\partial P_k} e^{- \theta_k s_k} f(\bm{g}) = 0, \label{cnd:P} \\
%    &\frac{\partial L}{\partial W_k} = 1 - \lambda_k \theta_k \mathbb{E}_{\bm{g}} \left\{\frac{\partial s_k}{\partial W_k} e^{- \theta_k s_k}\right\} = 0, \label{cnd:W} \\
%    &\eqref{con:Pmax}, \; \eqref{con:QLnr}. \nonumber
%\end{align}
Since problem \eqref{prob:RALag} is a functional optimization problem and the expectation $\mathbb{E}_{\bm{g}} \left\{\cdot\right\}$ is not with closed-form expression, neither analytical nor numerical solution of the problem can be found in general cases.

\subsection{Optimal Solution in Symmetric Scenario} \label{sec:Sym}
To provide a baseline for the learning-based solution to be introduced later, in what follows we find the optimal solution in a symmetric scenario, where all users are located at the cell-edge and have the same arrival process, i.e., $\alpha_k \!=\! \alpha$ and $a_k \!=\! a$. Then, $\theta_k \!=\! \theta$ and both the optimal values of $W_k$ and $\lambda_k$ are identical for different $k$, i.e., $W_k \!=\! W$ and $\lambda_k \!=\! \lambda$.

Denote the probability density function of $\bm{g}$ as $f(\bm{g})$. Then, the optimal solution of problem \eqref{prob:RALag} should satisfy its Karush-Kuhn-Tucker (KKT) conditions, which can be derived as,
\begin{align}
    &\frac{\mathrm{\delta} L}{\mathrm{\delta} P_k(\bm{g})} = h(\bm{g}) - \lambda \theta \frac{\partial s_k}{\partial P_k} e^{- \theta s_k} f(\bm{g}) = 0, \label{cnd:P} \\
    &\frac{\partial L}{\partial W} = 1 - \lambda \theta \mathbb{E}_{\bm{g}} \left\{\frac{\partial s_k}{\partial W} e^{- \theta s_k}\right\} = 0, \label{cnd:W} \\
    &\eqref{con:Pmax}, \; \eqref{con:QLnr}. \nonumber
\end{align}

\subsubsection{Optimal Power Allocation}
From \eqref{cnd:P} we have
\begin{align}   \label{eq:h}
    h(\bm{g}) &\!=\! \lambda \theta \frac{\partial s_k}{\partial P_k} e^{- \theta s_k} f(\bm{g}) \nonumber \\
    &\!=\! \lambda \theta \frac{\tau W}{u \ln{2}} \frac{\alpha g_k}{N_0 W} \frac{1}{(1 \!+\! \gamma_k)} e^{- \theta s_k} f(\bm{g}) \nonumber \\
    &\!=\! \frac{\lambda \theta \alpha g_k \tau}{N_0 u \ln\!{2} \left(1 \!+\! \gamma_k\right)} {\left(1 \!+\! \gamma_k\right)}^{-\frac{\theta W \tau}{u \ln\!{2}}} e^{\frac{\theta \sqrt{W \tau} Q_\mathrm{G}^{-1}\!\left({\varepsilon_\mathrm{max}/2}\right)}{u \ln\!{2}}} f(\bm{g}) \nonumber \\
    &\!=\! \frac{\beta g_k f(\bm{g})}{{\left(1 \!+\! \gamma_k\right)}^{\frac{1}{\eta}}},
\end{align}
where $\gamma_k \! \triangleq \! \frac{\alpha g_k P_k(\bm{g})}{N_0 W}$ is the SNR of the $k$th user, $\beta \! \triangleq \! \frac{\lambda \theta \alpha \tau}{N_0 u \ln\!{2}} e^{\frac{\theta \sqrt{W \tau} Q_\mathrm{G}^{-1}\!\left({\varepsilon_\mathrm{max}/2}\right)}{u \ln\!{2}}}$, and $\eta \! \triangleq \! 1/{\left({1+\frac{\theta W \tau}{u \ln\!{2}}}\right)}$.

Then, the power allocation function for the $k$th user can be derived from \eqref{eq:h} as
\begin{align}   \label{eq:P}
    P_k(\bm{g}) = \frac{N_0 W}{\alpha g_k} \left[{\left( \frac{\beta g_k f(\bm{g})}{h(\bm{g})} \right)}^\eta - 1\right].
\end{align}
Substituting \eqref{eq:P} into the equality in the maximum power constraint in \eqref{con:Pmax}, we have
\begin{align}   \label{opt:h}
    \sum_{k=1}^{K} {\frac{N_0 W}{\alpha g_k} \!\left[{\left( \frac{\beta g_k f(\bm{g})}{h(\bm{g})} \right)}^\eta \!-\! 1\right]} \!=& P_\mathrm{max}, \nonumber
\end{align}
from which we obtain
\begin{align}
%    {\left( \frac{\beta f(\bm{g})}{h(\bm{g})} \right)}^\eta \sum_{k=1}^{K} {{g_k}^{\eta-1}} \!-\! \sum_{k=1}^{K} {g_k^{-1}} \!=& \frac{\alpha P_\mathrm{max}}{N_0 W} \nonumber \\
    {\left( \frac{\beta f(\bm{g})}{h(\bm{g})} \right)}^\eta \!=& \frac{\frac{\alpha P_\mathrm{max}}{N_0 W} \!+\! \sum_{k=1}^{K} {{g_k}^{-1}}}{\sum_{k=1}^{K} {{g_k}^{\eta-1}}}.
\end{align}
Substituting \eqref{opt:h} into \eqref{eq:P}, the optimal power allocation function can be obtained as,
\begin{align}   \label{opt:P}
    P_k(\bm{g}) = \frac{N_0 W}{\alpha g_k} \left(\frac{\frac{\alpha g_k P_\mathrm{max}}{N_0 W} + g_k \sum_{i=1}^{K} {{g_i}^{-1}}}{{g_k}^{1-\eta} \sum_{i=1}^{K} {{g_i}^{\eta-1}}} - 1\right),
\end{align}
which does not depend on the channel distribution $f(\bm{g})$.

\subsubsection{Optimal Bandwidth Allocation}
With the optimal power allocation function, the optimal bandwidth allocated to each user can be found from the equality constraint in \eqref{con:QLnr}. Due to the expectation in \eqref{con:QLnr} and the complex expression of the achievable rate in \eqref{con:Srv}, the property of \eqref{con:QLnr} is hard to analyze. In concept, the achievable rate should increase with the bandwidth. However, this may not be true when the small-scale channel gain is very small (lower than $-10$~dB) due to the approximation $V_k \!\approx\! 1$. Fortunately, since very small values of the small-scale channel gain rarely occur (e.g., $\Pr\{g_k \!<\! 0.1\} \!<\! 10^{-12}$ when $N_\mathrm{t} \!\geq\! 8$ for Rayleigh fading channels), the impact can be ignored after taking the expectation. Therefore, it is reasonable to assume that the left-hand side of \eqref{con:QLnr} decreases with $W$. Then, the optimal bandwidth allocation can be found with stochastic optimization through the following iterations,
\begin{align}   \label{opt:W}
    W^{(t+1)} = {\left[W^{(t)} + \phi(t) \left(e^{- \theta s_k^{(t)}} - e^{-\theta B^\mathrm{E}}\right)\right]}^+,
\end{align}
where ${\left[x\right]}^+ \!\triangleq\! \max\!{\left\{x,0\right\}}$ ensures the results to be positive, $\phi(t) \!>\! 0$ is the step size, and $s_k^{(t)}$ is the achievable rate computed from the realization of $\bm{g}$ in the $t$th iteration. With the aforementioned assumption (which is true as we have validated via simulations) and $\phi(t) \!\sim\! \mathcal{O}\!\left(\frac{1}{t}\right)$, $\{W^{(t)}\}$ converges to the unique optimal bandwidth \cite{Bottou1998Stochastic}.

\begin{rem}
    \em{The KKT conditions are necessary for finding the global optimal solution. Since the power allocation derived from the KKT conditions and the bandwidth allocation found with stochastic optimization to satisfy the KKT condition are unique, the obtained solution is globally optimal.}
\end{rem}

\subsection{Solution with Unsupervised Learning  in General Case} \label{sec:DGM}
The difficulty in solving problem \eqref{prob:RALag} lies in finding the optimal power
allocation function $P_k(\bm{g})$, which does not have analytical expression in general case. Considering that neural networks are powerful at function learning,
%DGM. One of the advantages of the DGM, is that DGM does not suffer from the curse of dimension \cite{Sirignano2018DGM}. Such an advantage is crucial to the problem at hand, since $K$ power allocation functions need to be found and the number of users $K$ is usually large.
%\begin{figure}[htbp]
%	\vspace{-0.2cm}
%	\centering
%	\begin{minipage}[t]{0.48\textwidth}
%	\includegraphics[width=1\textwidth]{NN}
%	\end{minipage}
%	\vspace{-0.3cm}
%	\caption{The structure of the neural network for power allocation.}
%	\label{fig:NN}
%	\vspace{-0.2cm}
%\end{figure}
we approximate $P_k(\bm{g})$ with a parameterized function $\hat{P}_k(\bm{g};\bm{\omega})$, and
\begin{align}   \label{eq:NN}
    {\left[\hat{P}_1(\bm{g};\bm{\omega}),\cdots\!,\hat{P}_K(\bm{g};\bm{\omega})\right]}^\mathrm{T} = P_\mathrm{max} \mathcal{N}(\bm{g};\bm{\omega}),
\end{align}
where $\mathcal{N}(\bm{g};\bm{\omega})$ is a fully connected neural network with inputs $\bm{g}$ and parameters $\bm{\omega}$.

Then, we train $\bm{\omega}$ together with the bandwidth to obtain an approximated optimal resource allocation of the functional optimization problem. By applying \texttt{Softmax} in the output layer, $\hat{P}_k(\bm{g};\bm{\omega})$ automatically satisfies the maximum transmit power constraint. We use \texttt{ReLU} in the hidden layers as an example activation function, while similar results can be obtained with other activation functions. The width of each hidden layer is set as the number of users. By replacing $P_k(\bm{g})$ in \eqref{prob:RALag} with $\hat{P}_k(\bm{g};\bm{\omega})$, the optimization problem then becomes,
\begin{align}   \label{prob:RANN}
    \mathop \mathrm{min} \limits_{\bm{\omega}, W_k} \!\mathop \mathrm{max} \limits_{\lambda_k} \ & \hat{L} \! \triangleq\! \sum_{k=1}^{K} \left[{W_k} \!+\! {\lambda_k \!\left( \mathbb{E}_{\bm{g}} \!\left\{\!e^{- \theta_k \hat{s}_k}\!\right\} \!-\! e^{-\theta_k B^\mathrm{E}_k} \right)}\right] \\
    \text{s.t.} \ & \hat{s}_k \!=\! \frac{\tau W_k}{u \ln\!{2}} \left[\ln\!\left(\!1 \!+\! \frac{\alpha_k g_k \hat{P}_k(\bm{g};\bm{\omega})}{N_0 W_k}\!\right) \!-\! \frac{Q_\mathrm{G}^{-1}\!\left({\varepsilon^\mathrm{c}_k}\right)} {\sqrt{\tau W_k}}\right], \label{con:SrvApprox} \tag{\theequation a} \\
    & W_k \!\geq\! 0, \lambda_k \!\geq\! 0, \nonumber
\end{align}
where the term corresponds to the maximum transmit power constraint is omitted in the objective function since it is always zero for $\hat{P}_k(\bm{g};\bm{\omega})$.

By taking the Lagrange function $\hat{L}$ as the loss function, we can use SGD to find $\bm{\omega}$, $W_k$ and $\lambda_k$ in the following way,
\begin{align}
    \bm{\omega}^{(t+1)} &=\! \bm{\omega}^{(t)} \!-\! \phi(t) \nabla_{\bm{\omega}} \hat{L}^{(t)} \nonumber \\
    &=\! \bm{\omega}^{(t)} \!-\! \phi(t) P_\mathrm{max} \nabla_{\bm{\omega}} \mathcal{N}\left(\bm{g};\bm{\omega}^{(t)}\right) \nabla_{\hat{\bm{P}}} \hat{L}^{(t)}, \label{trn:Para} \\
    W_k^{(t+1)} &=\! {\left[W_k^{(t)} \!-\! \phi(t) \frac{\partial \hat{L}^{(t)}}{\partial W_k}\right]}^+, \label{trn:BW} \\
    \lambda_k^{(t+1)} &=\! {\left[\lambda_k^{(t)} \!+\! \phi(t) \frac{\partial \hat{L}^{(t)}}{\partial \lambda_k}\right]}^+ \nonumber \\
    &=\! {\left[\lambda_k^{(t)} \!+\! \phi(t) \frac{1}{N_\mathrm{b}} \sum_{n=1}^{N_\mathrm{b}}{\left( e^{-\! \theta_k \hat{s}_{k,n}^{(t)}} \!\!-\! e^{-\theta_k B^\mathrm{E}_k} \right)}\right]}^+, \label{trn:Lag}
\end{align}
where $\hat{L}^{(t)} \!\triangleq\! \frac{1}{N_\mathrm{b}} \sum_{n=1}^{N_\mathrm{b}}\sum_{k=1}^{K} {\left[W_k \!+\! \lambda_k \!\left( e^{- \theta_k \hat{s}_{k,n}^{(t)}} \!-\! e^{-\theta_k B^\mathrm{E}_k} \right)\right]}$, $\hat{s}_{k,n}^{(t)}$ is the $n$th realization of the achievable rate in the $t$th iteration, and $N_\mathrm{b}$ is the batch size in each iteration.
The gradient matrix of the neural network with respect to the parameters $\nabla_{\bm{\omega}} \mathcal{N}\left(\bm{g};\bm{\omega}^{(t)}\right)$ can be computed through backward propagation, and the gradient $\nabla_{\hat{\bm{P}}} \hat{L}^{(t)}$ is a column vector consists of $\left\{ -\! \frac{1}{N_\mathrm{b}} \sum_{n=1}^{N_\mathrm{b}} {\lambda_k^{(t)} \theta_k \frac{\partial \hat{s}_{k,n}^{(t)}}{\partial \hat{P}_k} e^{- \theta_k \hat{s}_{k,n}^{(t)}}} \bigg| k \!=\! 1,\cdots\!,K \right\}$.

\begin{rem}
\em{From the iteration of the Lagrange multiplier in \eqref{trn:Lag}, we can find that the iteration converges only when the QoS constraint \eqref{con:QLnr} is satisfied. This means that the QoS requirements can be ensured when the iteration converges.}
\end{rem}
\begin{rem}
    \em{The loss function $\hat{L}$ does not include any labels required in supervised learning. Hence, the way we employed to solve problem \eqref{prob:RANN} (and hence problem \eqref{prob:RA}) is unsupervised learning. We can observe from the iteration formulas that the property that the optimal solution should satisfy (i.e., the KKT conditions) is used as the ``supervision signal'' implicitly.}
\end{rem}

\section{Simulation Results}    \label{sec:Results}
In this section, we evaluate the minimal total bandwidth required to ensure the QoS by the optimal resource allocation, the approximated optimal solution and existing policies via simulations in both symmetric and asymmetric scenarios.

The cell radius is 250 m. In the symmetric scenario, all users are in the cell-edge. In the asymmetric scenario, the users are uniformly located in a road, where the user-BS distances are from $50$~m to $250$~m. The small scale channel gains of all users in each frame are randomly generated from Rayleigh distribution, and are independent from those in other frames. Other simulation parameters and fine-tuned hyper-parameters for the neural network are listed in Table \ref{tab:SimParam}.

\begin{table}[htbp]
	%\vspace{-0.2cm}
	\small
	\renewcommand{\arraystretch}{1.3}
	\caption{Simulation Parameters and Hyper-parameters}	\label{tab:SimParam}
	\begin{center}\vspace{-0.4cm}
	\begin{tabular}{|p{5.0cm}|p{2.9cm}|}
		\hline
        Overall packet loss probability $\varepsilon_\mathrm{max}$ & $10^{-5}$ \\ \hline
		Duration of each frame $T_\mathrm{f}$ & $0.1$~ms \\ \hline
		Duration of DL transmission  $\tau$ & $0.05$~ms \\ \hline
        DL delay bound $D_\mathrm{max}$ & $10$~frames ($1$~ms) \\ \hline
        Transmission delay $D^\mathrm{t}$ & $1$~frame \cite{Condoluci2017Reserv}\\ \hline
        Decoding delay $D^\mathrm{c}$ & $1$~frame \cite{Condoluci2017Reserv}\\ \hline
        Maximal transmit power of BS $P_\mathrm{max}$ & $43$~dBm \\ \hline
		Path loss model $10\lg(\alpha)$ & $35.3+37.6 \lg(d_k)$ \\ \hline
        Number of antennas $N_\mathrm{t}$ & 8 \\ \hline
		Single-sided noise spectral density $N_0$ & $-173$~dBm/Hz \\ \hline
		Packet size $u$ & $20$~bytes ($160$~bits) \cite{3GPP2016Scenarios} \\ \hline
        Average packet arrival rate $a$ & $0.2$~packets/frame \\ \hline
        Learning rate $\phi(t)$ & $1/(1+0.1 t)$ \\ \hline
        Number of hidden layers & $2$ \\ \hline
        Batch size $N_\mathrm{b}$ & $100$ \\ \hline
        % Iterations in each frame & $10$  \\ \hline
        %Total iteration steps & $1000$  \\ \hline
	\end{tabular}
	\end{center}
	\vspace{-0.7cm}
\end{table}

%\begin{figure}[htbp]
%	\vspace{-0.2cm}
%	\centering
%	\begin{minipage}[t]{0.48\textwidth}
%	\includegraphics[width=1\textwidth]{PA}
%	\end{minipage}
%	\vspace{-0.3cm}
%	\caption{Optimal power allocated to one of the two users, $W \!=\! 0.5$~MHz and $\theta \!=\! 2$.}
%	\label{fig:PA}
%	\vspace{-0.2cm}
%\end{figure}
%We first consider the symmetric scenario in Section \ref{sec:Sym} where $K=2$. The user-BS distance is $250$~m. The optimal normalized power allocated to the $1$st user, $P_1/P_\mathrm{max}$, is shown as the surface in Fig. \ref{fig:PA}, which is computed with \eqref{opt:P}. The transmit power of the other user can be obtained as $P_2/P_\mathrm{max} \!=\! 1 \!-\! P_1/P_\mathrm{max}$, which is not shown in the figure. When the channels of both users are identical, each user is allocated half of the maximum transmit power. When the channels are different, the user with better channel is allocated less transmit power and the user with worse channel is allocated more. It illustrates that the mechanism of multi-user diversity is to allocate less resources to the users with better channel conditions since they require less resources to ensure the QoS. Then, more resources can be employed to improve the QoS for the users in deep fading. Such a multi-user diversity is very different from traditional one, which works on the opposite: allocating more resources to the users with better channel conditions.

 The results of the optimal policy are obtained from \eqref{opt:P} and around $200$ iterations from \eqref{opt:W} only in the symmetric scenario (with legend ``Opt. Policy").

 The results of the approximated optimal solution with learning are obtained from the iterations in \eqref{trn:Para}, \eqref{trn:BW} and \eqref{trn:Lag} with random initial values (with legend ``Approx. Policy"). In each frame, the channel realizations in the recent $N_\mathrm{b}$ frames are taken as a batch, which is used for $10$ iterations. The training procedure converges after $100$ frames, unless otherwise specified.

For comparison, we provide the results for Policy B in \cite{Chengjian2017GCw}, which is a heuristic policy that exploits multi-user diversity by scheduling the users according to the small-scale channel gains of users (with legend ``Heur. Policy"). %When the maximal power is unable to support the desired service rate and the decoding error probability of all packets, the packets for some users that are in deep fading will be proactively dropped to ensure the reliability of other packets.
We also provide the results for the policy optimized in \cite{She2018CrossLayer}, which does not exploit multi-user diversity (with legend ``no MU diversity").
%Both the transmit power and the bandwidth are allocated based on the large-scale channel gains of each user, and hence the resources cannot be shared among users according to their diverse small-scale channel gains.

%\begin{figure}[htbp]
%	\vspace{-0.45cm}
%	\centering
%	\begin{minipage}[t]{0.5\textwidth}
%	\subfigure[{Symmetric scenario.}]{\label{fig:BW_Sym}
%		\includegraphics[width=0.4942\textwidth]{BW_Sym}}
%    \vspace{0.2cm}
%	\subfigure[{Asymmetric scenario.}]{\label{fig:BW_Asym}
%		\includegraphics[width=0.4658\textwidth]{BW_Asym}}
%	\end{minipage}
%    \vspace{-0.4cm}
%    \caption{Total bandwidth required to support the QoS of each user.}
%	\vspace{-0.2cm}
%\end{figure}

\begin{figure}[htbp]
	\vspace{0.05cm}
	\centering
	\begin{minipage}[t]{0.48\textwidth}
	\subfigure[{Symmetric scenario.}]{\label{fig:BW_Sym}
		\includegraphics[width=0.507\textwidth]{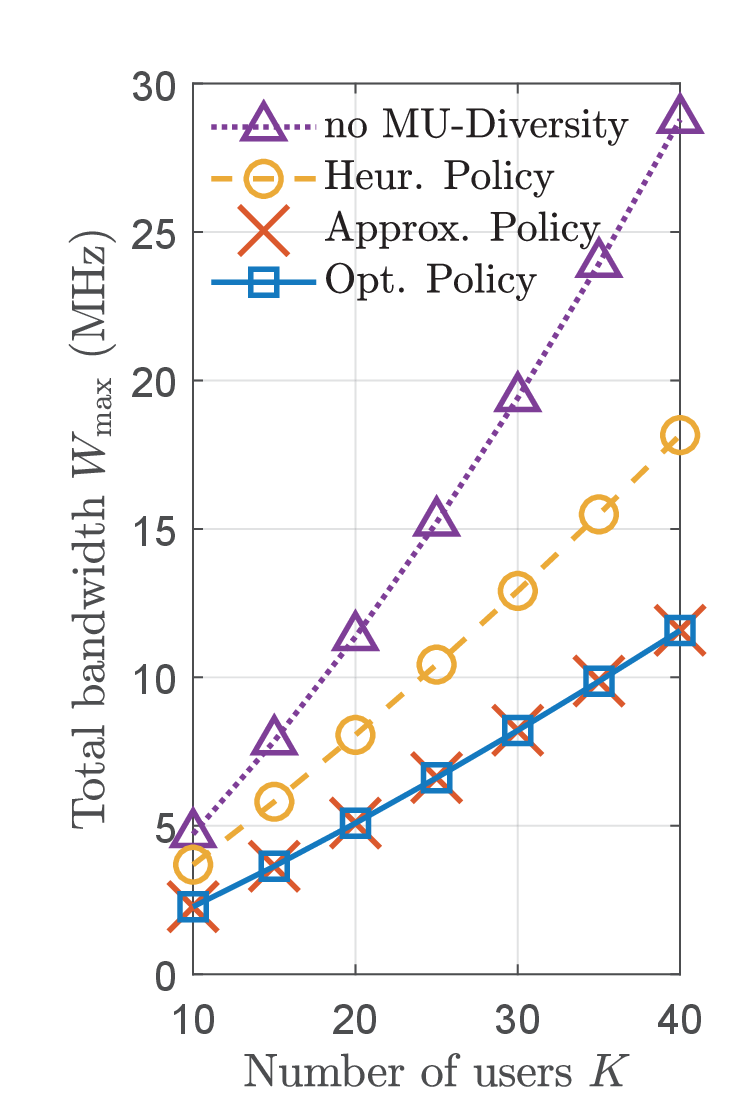}}
	\subfigure[{Asymmetric scenario.}]{\label{fig:BW_Asym}
		\includegraphics[width=0.453\textwidth]{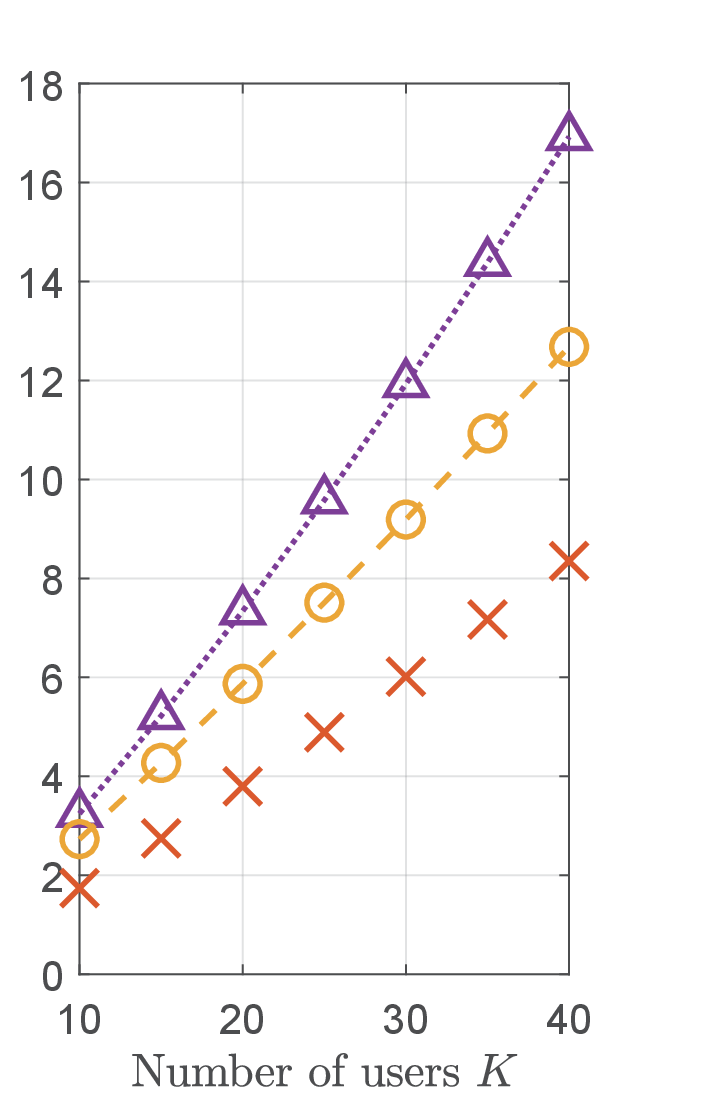}}
	\end{minipage}
    \vspace{-0.3cm}
    \caption{Total bandwidth required to support the QoS of each user.}
	\vspace{-0.2cm}
\end{figure}

 In Fig. \ref{fig:BW_Sym}, we provide the results in the symmetric scenario. It shows that the learning-based solution performs the same as the optimal policy, which means that the approximation is very accurate. Both policies can save about $60\%$ and $40\%$ of bandwidth compared with the policy without exploiting multi-user diversity and the heuristic policy, respectively. It is worthy to note that although the learning-based solution achieves optimal performance in this scenario, the symmetry assumption used in deriving the optimal solution is not employed during the training of the unsupervised learning.

 In Fig. \ref{fig:BW_Asym}, we  provide the results in the asymmetric scenario. It shows that the learning-based solution yields similar performance gain as in the symmetric scenario over the existing policies.
 %These results demonstrate that the learning-based solution can achieve the potential of  multi-user diversity.

\begin{table}[htbp]
	%\vspace{-0.2cm}
	\small
	\renewcommand{\arraystretch}{1.3}
	\caption{Number of Frames for Convergence (Asymmetric Scenario)}	\label{tab:CvgRate}
	\begin{center} \vspace{-0.4cm}
    \begin{tabular}{|c|c|c|}
    \hline
    Convergence percentage & $99.9\%$ & $99.99\%$ \\ \hline
    w/o pre-training & $5\,000$ & $>\!10\,000$ \\ \hline
    with pre-training & $3$ & $1\,000$ \\ \hline
    \end{tabular}
    \end{center}
	\vspace{-0.4cm}
\end{table}

To show the convergency of the learning-based solution, we consider the absolute sum of the average gradients
$\zeta^{(t)} \!\triangleq\! {\left\|{\mathbb{E}_{\bm{g}} \!\left\{\nabla_{\bm{\omega}} \hat{L}^{(t)}\!\right\}}\right\|}_1 \!+\! \sum_{k=1}^{K} \!{\left|\mathbb{E}_{\bm{g}} \!\left\{\frac{\partial \hat{L}^{(t)}}{\partial W_k}\!\right\}\!\right|} \!+\! \sum_{k=1}^{K} \!{\left|\mathbb{E}_{\bm{g}} \!\left\{\frac{\partial \hat{L}^{(t)}}{\partial \lambda_k}\!\right\}\!\right|}$
and the average relative error of the QoS constraint
$\xi^{(t)} \!\triangleq\! \sum_{k=1}^{K} \!{{\left[\mathbb{E}_{\bm{g}} \!\left\{e^{\theta_k \left(B^\mathrm{E}_k - \hat{s}_k^{(t)}\right)}\!\!\right\} \!-\! 1\right]}^+} \!\!\Big/\! K$.
The training algorithm in \eqref{trn:Para}, \eqref{trn:BW} and \eqref{trn:Lag} is considered to be converged at the $t$th frame when $\zeta^{(t)} \!<\! 1\% \!\times\! \sum_{k=1}^{K} {W_k^{(t)}}$ and $\xi^{(t)} \!<\! 1\%$.

The convergence speeds with and without pre-training are shown in Table \ref{tab:CvgRate}, which are obtained from $100\,000$ simulations. In each simulation, $40$ users are randomly dropped on the road. For the results without pre-training, the parameters are trained with random initial values until convergence, which needs $10\,000$ frames (i.e., 1 s) for 99.99\% convergence. For the results with pre-training, all users move at $72$~kph along the road in the same direction, and the parameters are retrained every 0.1 s by taking the pre-trained parameters as the initial values.
We can see that the pre-training can significantly shorten the convergence time, which can be done off-line.

The complexity of the training algorithm is low. A computer with {Intel\textregistered}\ {Core\texttrademark}\ {i7-6700} CPU is able to finish around $1\,000$ iterations in $0.1$~s without using the acceleration from GPU.
%Therefore, with pre-training, the algorithm can be efficiently carried out online.

\section{Conclusion}
In this paper, we proposed an approach of using un-supervised deep learning to solve the functional optimization problems with constraints. We considered an example problem of exploiting multi-user diversity in URLLC, which jointly optimizes power and bandwidth allocation that minimizes the total bandwidth required to ensure the QoS of each user. The global optimal solution was obtained in a symmetric scenario. An unsupervised learning method with neural network was introduced to find the approximated optimal solution for general cases, where the KKT conditions are implicitly served as
the ``supervision signal''.
Simulation results showed that the learning-based solution can achieve the same performance with the optimal solution in the symmetric scenario and outperforms existing policies with or without multi-user diversity in both symmetric and general scenarios. The training algorithm is with low computational complexity and converges rapidly with pre-training.

\appendices

\bibliographystyle{IEEEtran}
\bibliography{ref}

\end{document}